\documentclass[conference,letterpaper]{IEEEtran}

\usepackage[utf8]{inputenc} 
\usepackage[T1]{fontenc}
\usepackage{url}
\usepackage{ifthen}
\usepackage{cite}
\usepackage[cmex10]{amsmath} 

\usepackage{xcolor}
\usepackage{amssymb}
\usepackage{ntheorem}
\usepackage{dsfont}
\usepackage{algorithm}
\usepackage{algorithmic}

\newtheorem{theorem}{Theorem}[section]
\newtheorem{corollary}{Corollary}[section]
\newtheorem{lemma}{Lemma}[section]
\newtheorem{definition}{Definition}[section]
\newtheorem{remark}{Remark}[section]
\numberwithin{equation}{section}

\newcommand{\E} {\mathbb{E}}

\renewcommand{\P} {\mathbb{P}}

\DeclareMathOperator*{\argmax}{arg\,max}

\newcommand{\beq}{ \begin{equation} }
	\newcommand{\eeq}{ \end{equation} }

\begin{document}

\title{Exact Graph Matching in Correlated Gaussian-Attributed Erd\H{o}s-R\'enyi Model} 

 \author{%
   \IEEEauthorblockN{Joonhyuk Yang and Hye Won Chung}
   \IEEEauthorblockA{School of Electrical Engineering\\
                     KAIST\\
                      \{joonhyuk.yang, hwchung\}@kaist.ac.kr}
                   }



\maketitle

\begin{abstract}
Graph matching problem aims to identify node correspondence between two or more correlated graphs. Previous studies have primarily focused on models where only edge information is provided. However, in many social networks, not only the relationships between users, represented by edges, but also their personal information, represented by features, are present. In this paper, we address the challenge of identifying node correspondence in correlated graphs, where additional node features exist, as in many real-world settings.
We propose a two-step procedure, where we initially match a subset of nodes only using edge information, and then match the remaining nodes using node features. We derive information-theoretic limits for exact graph matching on this model. Our approach provides a comprehensive solution to the real-world graph matching problem by providing systematic ways to utilize both edge and node information for exact matching of the graphs.
\end{abstract}
\section{Introduction}
Graph matching is a fundamental problem in the field of network analysis and has wide applications in the several areas including bioinformatics \cite{KHGP16}, pattern recognition \cite{BBM05}, and social network analysis \cite{NS09,KL13}. The goal of graph matching is to identify node correspondence between two graphs sharing a common set of nodes. However, real-world scenarios often present challenges where edges in two graphs do not precisely overlap. To resolve this issue, researchers have explored this problem under random graph models.  One prominent model in this context is the correlated Erdős-Rényi (ER) model, first proposed by Pedarsani and Grossglauser \cite{PG11}. The correlated Erdős-Rényi model exhibits edge correlation between two graphs with unknown node correspondence.

One of the fundamental problems in graph matching is to identify regimes where the exact matching of the nodes between two graphs is feasible/infeasible. There has been a long line of research on finding the information-theoretic limits for exact graph matching in the correlated ER model  \cite{CK16,CK17,WXY22}. There also have been some recent works on finding information-theoretic limits for exact graph matching in correlated Stochastic Block Models (SBMs)  \cite{OGE16,RS21,YC23}, where the graph has communities in which the nodes are more densely connected than the nodes across the communities.

There also exist some examples of social networks where user information is embedded not in the form of edges between users but as attributes of each node. For example, if a person subscribes to two different movie-watching websites such as Netflix and IMDB, this user may provide his/her own information including age, preferred genre, location, etc, as attributes on both sites, which may not be exactly the same but correlated. This type of user-dependent attributes can be used for recovering node (user) correspondence from the two different sources, and there have been recent works to identify information-theoretic limits \cite{DCK19,CMK18} for exact matching from the correlated attributes. 

In this work, we consider the exact graph matching where the user information is embedded both in the form of edges between users and the node attributes. This model reflects practical social networks such as Facebook or Twitter, where there exist node information in the form of node attributes as well as edge information (friendship/following) between users. To examine the graph matching problem using such a combined form of node information, we introduce a model called correlated Gaussian-attributed  Erd\H{o}s-R\'enyi graphs, where we have access to features of each node represented as correlated Gaussian vectors in addition to the correlated  Erd\H{o}s-R\'enyi graphs. We derive information-theoretic limits for exact matching on this model in terms of the correlation between the feature vectors and that of edges between the two graphs. 
In particular, our results reveal the interesting regime when exact matching is infeasible using only the edge information but becomes feasible when combined with the node feature information. This finding unveils the quantitative gain from the combination of two types of information for graph matching.

\subsection{Correlated Gaussian-attributed Erd\H{o}s-R\'enyi model}\label{sec:model}
We introduce the correlated Gaussian-attributed Erd\H{o}s-R\'enyi (ER) model. First, we generate two correlated Erd\H{o}s-R\'enyi graphs $G_1$ and $G'_2$ with the same vertex set $[n]:=\{1,\dots,n\}$ in the following way: For $(i,j) \in [n]\times [n]$, $i\neq j$,
\begin{itemize}
    \item with probability $p_{11},$ it becomes an edge in both graphs;
    \item with probability $p_{10},$ it becomes an edge only in $G_1$;
    \item with probability $p_{01},$ it becomes an edge only in $G'_2$;
    \item with probability $p_{00}$, it remains unconnected in both graphs,
\end{itemize}
where $p_{11}+p_{10}+p_{01}+p_{00}=1$. 

Let us denote $V_1$ as the vertex set of $G_1$ and $V_2$ as the vertex set of $G'_2$. Then, a $d$-dimension Gaussian feature vector is assigned to each node in $G_1$ and $G'_2$ in the following way:
\begin{itemize}
    \item For $i\in V_1$, we assign $x_i\sim \mathcal{N}(\vec{0},I_d)$ to vertex $i$;
    \item For $i\in V_2$, we assign $y_i=\rho x_i + \sqrt{1-\rho^2} z_i$ to vertex $i$, for some $\rho\in[0,1]$ where $z_i\sim \mathcal{N}(\vec{0},I_d)$.
\end{itemize}
This can be expressed differently as follows: for $i \in [n]$,
\begin{equation}
    (x_i,y_i)\sim \mathcal{N}\left(\vec{0},\Sigma_d \right), \text{ where }
\end{equation}
\begin{equation}
    \Sigma_d:= \left[
\begin{matrix}
    I_d & \text{diag}(\rho) \\
\text{diag}(\rho) & I_d \\
\end{matrix}
\right].
\end{equation}
We can refer to the features assigned to each node as a database and represent the databases of graphs $G_1$ and $G'_2$  as matrices $X:=[x_1, x_2, \ldots ,x_n]^\top \in \mathbb{R}^{n\times d}$ and  $Y':=[y_1, y_2, \ldots ,y_n]^\top \in \mathbb{R}^{n\times d}$, respectively.

The graph $G_2$ is obtained by permuting the vertices of graph $G'_2$ by a permutation $\pi_*  :  [n]\to [n]$.  Let $Y=[y_{\pi_*(1)},y_{\pi_*(2)},\ldots,y_{\pi_*(n)}]^\top \in \mathbb{R}^{n\times d}$ represent the database matrix for graph $G_2$. For simplicity, we write the pair $(G_1,G_2)$ generated by this method as $(G_1,G_2)\sim \mathcal{CGD}(n,\mathbf{p},d,\rho)$, where $\mathbf{p}=(p_{11},p_{10},p_{01},p_{00})$. We refer to the model with $\mathbf{p} = (0, 0, 0, 1)$, i.e., only having Gaussian features but not the edges, as the correlated Gaussian databases.

We can see that the graph structures of $G_1$ and $G_2$ exhibit edge correlation, while the features have node correlation. Our goal is to find the regime of parameters $(n,\mathbf{p},d,\rho)$ where there exists an estimator $\hat{\pi}$ that can exactly recover the permutation $\pi_*$ by observing the pair $(G_1, G_2)$. 



\subsection{Related work and our contribution}
In the correlated Erd\H{o}s-R\'enyi model, Cullina and Kiyavash \cite{CK16,CK17} showed that exact graph matching is possible if $np_{11}\geq \log n + \omega(1)$ with some additional conditions or $n(\sqrt{p_{11}p_{00}}-\sqrt{p_{10}p_{01}})^2 \geq 2 \log n + \omega(1)$,  and it is impossible if $np_{11}\leq \log n -\omega(1).$ Moreover, in the subsampling model where $G_1$ and $G_2'$ are obtained by independently subsampling the edges of a parent Erd\H{o}s-R\'enyi graph $G\sim \mathcal{G}(n,p)$ with probability $s$, i.e., $p_{11}=ps^2,\; p_{10}=p_{01}=ps(1-s),\;p_{00}=1-2ps+ps^2$, which is a special case of the correlated Erd\H{o}s-R\'enyi model, Wu et al.\cite{WXY22} showed that exact matching is possible if $n(\sqrt{p_{11}p_{00}}-\sqrt{p_{10}p_{01}})^2 \geq (1+\epsilon)\log n $ and impossible if $n(\sqrt{p_{11}p_{00}}-\sqrt{p_{10}p_{01}})^2 \leq (1-\epsilon) \log n$ for any $\epsilon>0$.

Zhang et al. \cite{ZWW21} proposed attributed Erd\H{o}s-R\'enyi pair $\mathcal{G}(n,\mathbf{p},m,\mathbf{q})$ where $\mathbf{p}=(p_{11},p_{10},p_{01},p_{00})$ and $\mathbf{q}=(q_{11},q_{10},q_{01},q_{00})$. 
This model consists of $n$ user nodes and $m$ attribute nodes. The generating edge probability between user nodes depends on $\mathbf{p}$, while the generating edge probability between user nodes and attribute nodes depends on $\mathbf{q}$. 
In this model, the authors addressed the problem of matching $n$ user nodes when the attribute pairs are given. They showed that exact graph matching is possible when $np_{11}+m(\sqrt{q_{11}q_{00}}-\sqrt{q_{10}q_{01}})^2 \geq \log n +\omega(1)$ and impossible when $np_{11}+mq_{11}\leq \log n -\omega(1)$. 
It can be found that exact matching becomes feasible in a wider regime through attribute nodes, i.e., when $\mathbf{q}\neq \mathbf{0}$. 

When only Gaussian features are assigned to each node, Dai et al. \cite{DCK19} showed that exact matching is possible if $\frac{d}{4} \log \frac{1}{1-\rho^2} \geq \log n+ \omega(1)$ and impossible if $\frac{d}{4} \log \frac{1}{1-\rho^2} \leq (1-\epsilon)\log n$ for any positive constant $\epsilon$.

We consider the correlated Gaussian-attributed Erd\H{o}s-R\'enyi model where attributes are given as correlated Gaussian features in addition to the correlated Erd\H{o}s-R\'enyi  graphs. In this model, we derive information-theoretic limits for exact graph matching.
In order to demonstrate the feasibility conditions, we extend the $k$-core matching, which explores the matchings with a minimum degree of $k$ in an intersection graph. The $k$-core matching has been extensively studied in \cite{CKNP20,GRS22,RS23}, but we extend the analysis to a more general edge probability $\mathbf{p}$ by selecting an appropriate $k$, thereby obtaining a desired matching size. Subsequently, for the remaining unmatched nodes, we perform matching using Gaussian attributes. 
Conversely, we analyze the optimal maximum a posteriori (MAP) estimator to find the infeasibility conditions for exact matching.

\subsection{Notation}
For any positive integer $n$, let $[n]$ be defined as $\{1, 2,\ldots, n\}$. For a graph $G$ with a vertex set $[n]$, let $\deg_G(i)$ denote the number of neighbors of $i\in [n]$ in $G$ and let $G\{M\}$ denote the subgraph of $G$ induced by $M\subset [n]$. Let $d_{\min}(G)$ be the minimum degree of a graph $G$. Let $\mathcal{E}:=\{ \{i,j\} : i,j\in [n], i \neq j \}$ denote the set of all unordered vertex pairs. Let $A$, $B'$, and $B$ be the adjacency matrices of $ G_1$, $G'_2$, and $G_2$, respectively. 
Asymptotic dependencies are denoted with standard notations $O(\cdot), o(\cdot), \Omega(\cdot), \omega(\cdot), \Theta(\cdot)$ with $n\to\infty$.  Let $\vee$ denote the max operator. 
For an event $E$, let $\mathds{1}(E)$ be the indicator random variable.

\section{Main Result}
Let $p_{1*}=p_{11}+p_{10},\; p_{*1}=p_{11}+p_{01}$. Our main results are as follows:
\begin{theorem}[Achievability]\label{thm:achievability}
    Consider the correlated Gaussian-attributed ER model $(G_1,G_2)\sim \mathcal{CGD}(n,\mathbf{p},d,\rho)$. For any arbitrary small constant $\epsilon >0$, if
    \begin{equation}\label{eq:achievability}
        np_{11} + \frac{d}{4} \log \frac{1}{1-\rho^2} \geq (1+\epsilon) \log n \text{ and}
    \end{equation}
            \begin{equation}\label{eq:achievability p regmie}
    \frac{p_{1*}p_{*1}}{p_{11}}\leq O\left(\frac{1}{e^{(\log \log n)^3}}\right),
\end{equation}
 then there exists an estimator $\hat{\pi}$ such that $\hat{\pi}=\pi_*$ with probability $1-o(1)$.
\end{theorem}

\begin{remark}
    In the subsampling model, the condition \eqref{eq:achievability p regmie} reads as $p \leq O\left(\frac{1}{e^{(\log \log n)^3}}\right)$. We can see that the edge probability $p$ of a parent graph can cover not only sparse regime $p=n^{-\Omega(1)}$, but also some of dense regime $p=n^{-o(1)}$.
\end{remark}

\begin{theorem}[Impossibility]\label{thm:impossibbility}
   Consider the correlated Gaussian-attributed ER model $(G_1,G_2)\sim \mathcal{CGD}(n,\mathbf{p},d,\rho)$. For any arbitrary small constant $\epsilon >0$, if $1\ll d =O(\log n)$,
    \begin{equation}\label{eq:impossible}
        np_{11} + \frac{d}{4} \log \frac{1}{1-\rho^2} \leq (1-\epsilon) \log n, \text{ and}
    \end{equation}
    \begin{equation}\label{eq:positive correlation}
        p_{11}p_{00}>p_{10}p_{01}
    \end{equation}
        then for any estimator $\hat{\pi}$, we obtain that $\P(\hat{\pi}=\pi_*)=o(1)$.
\end{theorem}

\begin{remark}
    If $p_{11}>p_{1*}p_{*1}$ then we say that the graphs $(G_1,G_2)$ exhibit a positive correlation. Note that $p_{11}>p_{1*}p_{*1}$ is equivalent to $p_{11}p_{00}>p_{10}p_{01}$. We only consider the case of positive correlation.
\end{remark}


\section{Proof of Theorem \ref{thm:achievability}}
\subsection{Algorithm and outline of proof}

 \begin{algorithm}[h]
    \caption{Achievability}
    \label{alg:achiev}
\begin{algorithmic}[1]
  \REQUIRE{$(G_1,G_2)\sim \mathcal{CGD}(n,\mathbf{p},d,\rho)$ and $k=\frac{np_{11}}{(\log np_{11})^2}\vee \frac{\log n}{(\log \log n)^2}$}
    \ENSURE{$\widehat{\pi} : [n] \to [n]$}
    \STATE $(\widehat{M}_k,\widehat{\mu}_k) \leftarrow (\varnothing,\varnothing)$
    \FOR{matching $ (M,\mu)$}
    \IF{ $d_{\min}(G_1 \wedge_\mu G_2)\geq k$ and $|\widehat{M}_k|<|M|$}
    \STATE $(\widehat{M}_k,\widehat{\mu}_k)\leftarrow (M,\mu)$
    \ENDIF
    \ENDFOR
    \STATE $J\leftarrow [n]\backslash \widehat{M}_k$, $J'\leftarrow [n]\backslash \widehat{\mu}_k(\widehat{M}_k)$
    \STATE $\Tilde{\mu}_k =\argmax_{\mu : J\to J'} \P\left( \left.\pi_*\{J\}=\mu\{J\}  \right| X,Y,J,J' \right)$ 
    \RETURN $\widehat{\pi} = (\widehat{\mu}_k, \Tilde{\mu}_k) : [n]\to [n]$    
\end{algorithmic}
\end{algorithm}
To prove Theorem \ref{thm:achievability}, we proceed through two steps. We first use the $k$-core estimator (as in lines $1$-$6$ in Algorithm \ref{alg:achiev}), which uses only edge information, i.e., adjacency matrices $A$ and $B$. By choosing $k=\frac{np_{11}}{(\log np_{11})^2}\vee \frac{\log n}{(\log \log n)^2}$, we can get a sufficiently large matching with no mismatched node pairs. We will match the unmatched nodes using the maximum a posteriori estimator for node information, i.e., database matrices $X$ and $Y$ (as in lines $7$-$8$ in Algorithm \ref{alg:achiev}). 

\subsection{$k$-core matching}\label{sec:k-core}
In this subsection, we present the results for partial matching and exact matching when using the $k$-core estimator.
\begin{definition}[Matching]\label{def:matching}
    Consider two graphs $G_1$ and $G_2$. $(M,\mu)$ is a matching between $G_1$ and $G_2$ if $M\subset [n]$ and $\mu $ : $M \to [n]$ is injective. For a matching $(M,\mu)$, we define $\mu(M)$ as the image of $M$ under $\mu$, and $\mu\{M\}:=\{(i,\mu(i)) : i \in [M]\}$.
\end{definition}

Given graphs $G_1$ and $G_2$ with matching $(M,\mu)$, let us define the intersection graph $G_1 \wedge_{\mu} G_2$ as follows: 
 For $u,v\in M$, $(u,v)$ is an edge in $G_1 \wedge_{\mu} G_2$ if and only if $(u,v)$ is an edge in $G_1$ and $(\mu(u),\mu(v))$ is an edge in $G_2$. The $k$-core matching and $k$-core estimator were defined in \cite{CKNP20,GRS22,RS23}. For the completeness, we provide its definition here once again.
\begin{definition}[$k$-core matching and $k$-core estimator]\label{def:k-core estimator}
    Consider two graphs $G_1$ and $G_2$. A matching $(M,\mu)$ is a $k$-core matching if $d_{\min}(G_1\wedge_{\mu} G_2)\geq k$. Furthermore, the $k$-core estimator $(\widehat{M}_k,\widehat{\mu}_k)$ is the $k$-core matching that includes the largest nodes among all $k$-core matchings.
\end{definition}
By using the $k$-core estimator, we can achieve a partial matching with no mismatched node pairs, as stated in the following theorem. 
\begin{theorem}[Partial matching from $k$-core estimator]\label{thm:k-core matching}
    Consider the correlated Erd\H{o}s-R\'enyi model $(G_1,G_2)\sim \mathcal{CG}(n,\mathbf{p})$. Suppose that 
      \begin{equation}\label{eq:k-core p regmie}
    \frac{p_{1*}p_{*1}}{p_{11}}\leq O\left(\frac{1}{e^{(\log \log n)^3}}\right) \text{ and}
\end{equation}
      \begin{equation}\label{eq:k-core k regmie}
   k=\frac{np_{11}}{(\log np_{11})^2} \vee \frac{\log n}{(\log \log n)^2}.
\end{equation}
Then, the $k$-core estimator $(\widehat{M}_k,\widehat{\mu}_k)$ satisfies that 
\begin{equation}\label{eq:k-core size}
    |\widehat{M}_k|\geq n-n^{1-\frac{np_{11}}{\log n}+o(1)} \text{ and}
\end{equation}
    \begin{equation}\label{eq:k-core groundtruth}
    \widehat{\mu}_k\{\widehat{M}_k\}=\pi_*\{\widehat{M}_k\}
    \end{equation}
  with probability $1-o(1)$.
\end{theorem}

We can also derive sufficient conditions for exact matching.

\begin{theorem}[Exact matching from $k$-core estimator]\label{thm:exact matching}
    Consider the correlated Erd\H{o}s-R\'enyi model $(G_1,G_2)\sim \mathcal{CG}(n,\mathbf{p})$. Suppose that $\eqref{eq:k-core p regmie}$ and \eqref{eq:k-core k regmie} hold. Also, assume that
    \begin{equation}\label{eq:exact matching}
        np_{11}\geq (1+\epsilon) \log n
    \end{equation}
    for any arbitrary small constant $\epsilon>0$. Then, $\widehat{\mu}_{k}=\pi_*$ with probability $1-o(1)$.
\end{theorem}

We will provide the proofs for Thm. \ref{thm:k-core matching}--\ref{thm:exact matching} in Sec. \ref{sec:k-core}. 

\subsection{Database matching}

Dai et al. \cite{DCK19} identified achievability conditions for exact matching using only the correlated Gaussian databases.
\begin{theorem}[Theorem 1 in \cite{DCK19}]\label{thm:dai achieve}
Consider the correlated Gaussian databases $X,Y \in \mathbb{R}^{n\times d}$ defined in Section \ref{sec:model}. Suppose that 
\begin{equation}
    \frac{d}{4}\log \frac{1}{1-\rho^2} \geq \log n +\omega(1).
\end{equation}
Then, the MAP estimator can exactly recover $\pi_*$ with probability $1-o(1).$
\end{theorem}

\begin{IEEEproof}[Proof of Theorem \ref{thm:achievability}]
Let $k=\frac{np_{11}}{(\log np_{11})^2} \vee \frac{\log n}{(\log \log n)^2}.$ If $np_{11}\geq (1+\epsilon) \log n$, then the $k$-core estimator achieves the exact matching from Theorem \ref{thm:exact matching}. Now, let us consider the case where $np_{11}\leq (1+\epsilon) \log n$. By Theorem \ref{thm:k-core matching}, we obtain the $k$-core estimator $(\widehat{M}_k,\widehat{\mu}_k)$ satisfying \eqref{eq:k-core size} and $\eqref{eq:k-core groundtruth}$. Let $J=[n]\backslash \widehat{M}_k$. We have that 
\begin{equation}\label{eq:size J}
    |J|\leq n^{1-\frac{np_{11}}{\log n}+o(1)}
\end{equation}
since \eqref{eq:k-core size}. By assumption \eqref{eq:achievability} and \eqref{eq:size J}, we have that
\begin{equation}
\begin{aligned}
      \frac{d}{4} \log \frac{1}{1-\rho^2} &\geq (1+\epsilon) \log n -np_{11}\\
      &\geq  \log |J| + \omega(1)
\end{aligned}
\end{equation}
Thus, the exact matching is possible on vertex set $J$ by using only features due to Theorem \ref{thm:dai achieve}.\end{IEEEproof}

\section{Proof of Theorem \ref{thm:impossibbility}}
Before proving Theorem \ref{thm:impossibbility}, we present analysis on the posterior distribution for a permutation $\pi$ given correlated Erd\H{o}s-R\'enyi graphs.
For $a,b\in \{0,1\}$, let us define
\begin{equation}
    \begin{aligned}
        \mu_{ab}(\pi):=\sum_{(i,j) \in \mathcal{E}} \mathds{1}\{(A_{i,j},B_{\pi(i),\pi(j)})=(a,b) \}
    \end{aligned}
\end{equation}
and
\begin{equation}
      p_{ab}:=\P\left((A_{i,j},B_{\pi(i),\pi(j)})=(a,b)\right).
\end{equation}  
Then, we have that
\begin{equation}\label{eq:express mu}
    \begin{aligned}
        \mu_{00}(\pi)&=\sum_{(i,j)\in \mathcal{E}} (1-A_{i,j})(1-B_{\pi(i),\pi(j)})\\
        &=\sum_{(i,j)\in \mathcal{E}} (1-A_{i,j}-B_{\pi(i),\pi(j)})+\mu_{11}(\pi),\\
        \mu_{10}(\pi)&=\sum_{(i,j)\in \mathcal{E}} A_{i,j}(1-B_{\pi(i),\pi(j)})\\
        &=\sum_{(i,j)\in \mathcal{E}} A_{i,j} -\mu_{11}(\pi),\\
        \mu_{01}(\pi)&=\sum_{(i,j)\in \mathcal{E}} (1-A_{i,j})B_{\pi(i),\pi(j)}\\
        &=\sum_{(i,j)\in \mathcal{E}} B_{\pi(i),\pi(j)} -\mu_{11}(\pi).
    \end{aligned}
\end{equation}
By \eqref{eq:express mu}, we obtain that 
\begin{equation}\label{eq:posterior}
    \begin{aligned}
        \P\left(\pi_{*}=\pi \mid A,B \right)&=\frac{ \P\left(A,B \mid \pi_*=\pi \right)\P(\pi_*=\pi)}{\P(A,B)}
       \\&=\frac{\P(\pi_*=\pi)}{\P(A,B)}p^{\mu_{00}(\pi)}_{00} p^{\mu_{10}(\pi)}_{10} p^{\mu_{01}(\pi)}_{01} p^{\mu_{11}(\pi)}_{11}\\
        &=c(A,B) \left(\frac{p_{00}p_{11}}{p_{10}p_{01}}\right)^{\mu_{11}(\pi)},
    \end{aligned}
\end{equation}
where $c(A,B)$ depends only on $A$ and $B$. The last equality holds from \eqref{eq:express mu}. Define the vertex set
\begin{equation}\label{def:H}
\begin{aligned}
    \mathcal{H}(\pi):=\left\{ i\in[n] : \forall j\in [n],\; A_{i,j}B_{\pi(i),\pi(j)}=0 \right\}.
\end{aligned}
\end{equation}
For notational simplicity, let $\mathcal{H}_*:= \mathcal{H}(\pi_*)$. The size of $\mathcal{H}_*$ is as follows:
\begin{lemma}\label{lem:size of H}
   If $np_{11}\leq \log n - \omega(1),$  then it holds that $|\mathcal{H}_*|\geq \frac{1}{4}n^{1-\frac{np_{11}}{\log n}}$,
    with probability $1-o(1)$.
\end{lemma}

Define the permutation set  $\mathcal{T}_*$ as follow: 
\begin{definition}\label{def:T}
  A permutation  $\pi\in \mathcal{T}_*$ if and only if the following conditions hold: 
            \begin{itemize}
        \item $\pi(i)=\pi_*(i)$  \quad \ if $i\in  [n] \backslash \mathcal{H}_*$
        \item $\pi(i)=\pi_*(\rho(i))$  if $i \in \mathcal{H}_*$,
    \end{itemize}
    where $\rho$ is any permutation over $\mathcal{H}_*$.
\end{definition}

\begin{lemma}\label{lem:mu}
    For any $\pi \in \mathcal{T}_*$, we have $\mu_{11}(\pi)\geq \mu_{11}(\pi_*)$.
\end{lemma}



Dai et al. \cite{DCK19} identified impossibility conditions of exact matching in the correlated Gaussian databases.
\begin{theorem}[Theorem 2 in \cite{DCK19}]\label{thm:dai impossible}
Consider the correlated Gaussian databases $X,Y \in \mathbb{R}^{n\times d}$ defined in Section \ref{sec:model}. Suppose that $1\ll d =O(\log n)$ and 
\begin{equation}
    \frac{d}{4}\log \frac{1}{1-\rho^2} \leq (1-\epsilon)\log n
\end{equation}
for any arbitrary small constant $\epsilon>0$. Then, for any algorithm, the probability that the algorithm outputs the correct permutation $\pi_*$ is $o(1).$
\end{theorem}

\begin{IEEEproof}[Proof of Theorem \ref{thm:impossibbility}]
We consider the posterior distribution of $\pi$ given not only $G_1,G_2$ but also $\pi_*\{[n]\backslash \mathcal{H}_*\}$. Then, the maximum a posterior (MAP) estimator is given by
\begin{equation}\label{eq:MAP}
    \widehat{\pi}_{\text{MAP}}:=\argmax_{\pi}\P\left(\pi_{*}=\pi \mid G_1, G_2, \pi_*\{[n]\backslash \mathcal{H}_*\} \right).
\end{equation}
We obtain that
\begin{equation}\label{eq:posterior addi}
\begin{aligned}
    &\P\left(\pi_{*}=\pi \mid G_1, G_2, \pi_*\{[n]\backslash \mathcal{H}_*\} \right) \\
    &=\frac{\P(\pi_*=\pi | G_1,G_2)\P( \pi_*\{[n]\backslash \mathcal{H}_*\} | \pi_*=\pi,G_1,G_2)}{\P( \pi_*\{[n]\backslash \mathcal{H}_*\} |  G_1,G_2)}\\
    &\stackrel{(a)}{=}\frac{\P(\pi_*=\pi | A,B,X,Y)}{\P( \pi_*\{[n]\backslash \mathcal{H}_* \}|  A,B,X,Y)} \mathds{1}(\pi \in \mathcal{T}_*)\\
    &\stackrel{(b)}{=} \frac{\P(\pi_*=\pi | A,B)\P(\pi_*=\pi | X,Y)}{\P(\pi_*=\pi)\P(\pi_*\{[n]\backslash \mathcal{H}_* \}|  A,B,X,Y)} \mathds{1}(\pi \in \mathcal{T}_*)\\
    &=C_1 \P\left(\pi_{*}=\pi \mid A,B \right)\P\left(\pi_{*}=\pi \mid X,Y \right) \mathds{1}(\pi \in \mathcal{T}_*)\\
    &=C_2\P\left(\pi_{*}=\pi \mid X,Y \right) \left(\frac{p_{00}p_{11}}{p_{10}p_{01}}\right)^{\mu_{11}(\pi)}\mathds{1}(\pi \in \mathcal{T}_*)
\end{aligned}
\end{equation}
where $C_1,C_2$ are constants depending on $A,B, \pi_*\{[n]\backslash \mathcal{H}_* \}$.
The equality $(a)$ holds by Definition \ref{def:T}. The 
equality $(b)$ holds since adjacency matrices $A,B$ and databases $X,Y$ are independent when $\pi_*=\pi$ is given, and the last equality holds from \eqref{eq:posterior}.

Combining Lemma \ref{lem:mu} and the assumption \eqref{eq:positive correlation}, for $\pi \in \mathcal{T}_*$, we can obtain that
\begin{equation}
      \left(\frac{p_{00}p_{11}}{p_{10}p_{01}}\right)^{\mu_{11}(\pi)} \geq  \left(\frac{p_{00}p_{11}}{p_{10}p_{01}}\right)^{\mu_{11}(\pi_*)}.
\end{equation}
Furthermore, $|\mathcal{H}_*|=\omega(1)$. Thus, if we show that exact matching is impossible between the nodes in $\mathcal{H}_*$ by using database matrices $X,Y$, then it is possible to show the failure of the MAP estimator.
Combining Lemma \ref{lem:size of H} and assumption \eqref{eq:achievability}, we can obtain that 
\begin{equation}
\begin{aligned}
     \frac{d}{4} \log \frac{1}{1-\rho^2} &\leq (1-\epsilon) \log n -np_{11}\\
     &\leq \left(1-\frac{\epsilon}{2}\right) \log |\mathcal{H}_*|.
\end{aligned}
\end{equation}
Thus, the MAP estimator fails by Theorem \ref{thm:dai impossible}.
\end{IEEEproof}

To prove Theorem \ref{thm:impossibbility}, we utilized Theorem \ref{thm:dai impossible}. In order to apply Theorem \ref{thm:dai impossible}, the condition $1 \ll d = O(\log n)$ is necessary. Wang et al. \cite{Wang22} have obtained even tighter results. They showed that if $\frac{1}{\rho^2}-1 \leq d/40$ and $\frac{d}{4}\log \frac{1}{1-\rho^2}\leq \log n -\log d +C$ for a constant $C>0$, then there is no algorithm that guarantees exact matching with high probability.  Applying this result, we can obtain the following corollary.

\begin{corollary}
  Consider the correlated attributed Erd\H{o}s-R\'enyi model $(G_1,G_2)\sim \mathcal{CGD}(n,\mathbf{p},d,\rho)$. For any arbitrary small constant $\epsilon >0$, if $ p_{11}p_{00}>p_{10}p_{01}, \frac{1}{\rho^2}-1\leq d/40$ and
    \begin{equation}\label{eq:impossible}
        np_{11} + \frac{d}{4} \log \frac{1}{1-\rho^2} \leq \log n - \log d -\omega(1)
    \end{equation}
        then for any estimator $\hat{\pi}$, we obtain that $\P(\hat{\pi}=\pi_*)=o(1)$.    
\end{corollary}

\section{Analysis of $k$-core matching}\label{sec:k-core}
In this section, we introduce lemmas regarding the size and accuracy of matching obtained using the $k$-core estimator. For a graph $G$, define the set 
$
    L_k := \{i\in [n] : \deg_G(i) \leq k\}.
$
\begin{lemma}\label{lem:size degree k}
    Let $G\sim \mathcal{G}(n,p)$. Then, it holds that
    \begin{equation}
        \E[|L_k|]\leq n \exp(-np+k\log np+1).
    \end{equation}
\end{lemma}

\begin{definition}
    For a graph $G$, a vertex set $M$ is called the $k$-core of the graph $G$ if it is the largest set such that $d_{\min}(G\{M\})\geq k$.
\end{definition}

For the correlated  Erd\H{o}s-R\'enyi model $(G_1,G_2)\sim \mathcal{CG}(n,\mathbf{p})$, define the set $M_k$ as the largest set $M_k\subset [n]$ satisfying $d_{\min}(G_1 \wedge_{\pi_*}G_2\{M_k\}) \geq k$. We refer to $M_k$ as the $k$-core of the graph $G_1 \wedge_{\pi_*} G_2$. We provide the lemma for $k$-core matching, similar to \cite{CKNP20,GRS22,RS23}, but we extend the results by selecting a proper $k$ to obtain a large enough matching as well as to cover a dense regime.  In \cite{CKNP20}, Cullina et al. chose a large value of $k\geq \Omega(np_{11})$, resulting in too small matching size $|M _k|$. In \cite{GRS22,RS23}, the authors chose a constant $k$, resulting in the condition  $p_{11}+p_{10},p_{11}+p_{01} = o(n^{-1/2})$ to prevent the occurrence of mismatched node pairs. By selecting a proper $k$, we obtain a significant matching size without any mismatched node pairs, except in highly dense regime.
\color{black}

\begin{lemma}\label{lem:k-core}
     Consider the correlated Erd\H{o}s-R\'enyi model $(G_1,G_2)\sim \mathcal{CG}(n,\mathbf{p})$. Suppose that \eqref{eq:k-core p regmie} and \eqref{eq:k-core k regmie} hold. Then, $k$-core estimator $(\widehat{M}_k,\widehat{\mu}_k)$ satisfies that 
\begin{equation}
    \widehat{M}_k=M_k \text{ and } \widehat{\mu}_k\{\widehat{M}_k\}=\pi_*\{\widehat{M}_k\}
\end{equation}
  with probability $1-o(1)$.
\end{lemma}

\begin{IEEEproof}[Proof of Theorem \ref{thm:exact matching}]
    From Lemma \ref{lem:k-core}, we obtain that $ \widehat{M}_k=M_k$ and $\widehat{\mu}_k\{\widehat{M}_k\}=\pi_*\{\widehat{M}_k\}$. Thus, it suffices to prove that $|M_k|=n$ with high probability.
    We can obtain
    \begin{equation}
        \begin{aligned}
            \P(|M_k|\neq n)&=\P(d_{\min}(G_1\wedge_{\pi_*}G_2)<k)\\
            & \stackrel{(a)}{\leq} n\P(\deg_{G_1\wedge_{\pi_*}G_2}(i)<k)\\
            & \stackrel{(b)}{\leq} n \exp(-np_{11}+(k-1) \log np_{11}+1)\\
            & = o(1)            .
        \end{aligned}
    \end{equation}
    The inequality $(a)$ holds by taking union bound. The inequality $(b)$ holds by Lemma \ref{lem:size degree k}. The last equality holds due to assumptions \eqref{eq:k-core k regmie} and \eqref{eq:exact matching}. Hence, $|M_k|=n$ with probability $1-o(1)$ and the proof is complete.
\end{IEEEproof}

Define  the set
\begin{equation}\label{eq:Lhat}
    \widehat{L}_k := \{i\in [n] : \deg_{G_1 \wedge_{\pi_*}G_2}(i) \leq k\}.
\end{equation}

Let $J_k := [n] \backslash M_k$. We can obtain the following lemma similar to \cite{Luczak,RS23}.
    
\begin{lemma}\label{lem:3L}
    Consider the correlated Erd\H{o}s-R\'enyi model $(G_1,G_2)\sim \mathcal{CG}(n,\mathbf{p})$. 
    Suppose that $np_{11}=\Theta\left(\log n\right)$ and $\eqref{eq:k-core k regmie}$ holds. Then, it holds that $|J_k| \leq 3|\widehat{L}_{k+1}|$.
\end{lemma}

\begin{IEEEproof}[Proof of Theorem \ref{thm:k-core matching}]
If $np_{11}\geq (1+\epsilon) \log n$ for any arbitrary small constant $\epsilon>0$, then $\widehat{\mu}_k=\pi_*$ due to Theorem \ref{thm:exact matching}. Thus, we consider only $np_{11}=O(\log n)$. By Lemma \ref{lem:k-core}, we can have that $ \widehat{M}_k=M_k$ and $\widehat{\mu}_k\{\widehat{M}_k\}=\pi_*\{\widehat{M}_k\}$. Thus, it suffices to show that $|M_k|\geq n-n^{1-\frac{np_{11}}{\log n} + o(1)}$ with high probability. If $np_{11}=o(\log n)$, then it is trivial since $n-n^{1-\frac{np_{11}}{\log n} + o(1)} \leq 0$. Therefore, let us consider the case where $np_{11}=\Theta(\log n)$. 

Let $J_k = [n] \backslash M_k$ and recall $\widehat{L}_k$ defined in \eqref{eq:Lhat}
     By Markov's inequality, we have $
         |\widehat{L}_{k+1}|\leq (\log n)\E[|\widehat{L}_{k+1}|]$
     with probability at least $1-\frac{1}{\log n}$. Thus, it holds that
     \begin{equation}
         |J_k|\leq 3|\widehat{L}_{k+1}| \leq n^{1-\frac{np_{11}-(k+1) \log np_{11}}{\log n}+o(1)}
     \end{equation}
     due to Lemma \ref{lem:size degree k} and Lemma \ref{lem:3L}. By assumption $\eqref{eq:k-core k regmie}$, we have 
     $\frac{(k+1) \log np_{11}}{\log n} =o(1)$. Thus, the proof is complete.
\end{IEEEproof}
\section{Conclusion and open problems}
In this paper, we proposed a new model called the correlated Gaussian-attributed Erdős-Rényi model, where features for each node are given as correlated Gaussian vectors, in addition to correlated edges. We derived the information-theoretic limits for exact matching, which reveals the quantitative relationship between information on edges and on node features for exact matching.
Our work leaves interesting open problems:
\begin{itemize}
    \item \textbf{Finding a tighter information-theoretic limits :} The threshold for achievability of exact matching in both the correlated Erdős-Rényi model and Gaussian databases is $\log n + \omega(1)$. This raises the question of whether the condition for achievability in \eqref{eq:achievability} can be further tightened to  $np_{11}+\frac{d}{4}\log \frac{1}{1-\rho^2} \geq \log n +\omega(1)$ in the correlated Gaussian attributed Erdős-Rényi model, which is an open problem.
        \item \textbf{Finding efficient algorithm for exact matching :}  In the correlated Gaussian databases, the MAP estimator requires a time complexity of $O(n^3 + n^2 d)$ in computing the joint likelihood for all pairs of features and finding the permutation that maximizes the value through the Hungarian algorithm. However, the $k$-core matching requires considering all possible matchings with the time complexity of $\Theta(n!)$. Designing a polynomial-time algorithm achieving the information-theoretic limit remains as an open problem. In the correlated Erdős-Rényi model, various efficient algorithms have been proposed, including seeded \cite{YG13, MX20} and seedless matching \cite{DMWX21, FMWX23, MRT23, MWXY23}. In the correlated SBMs, Yang et al. \cite{YSC23} proposed an efficient algorithm by using community structure, which is a type of attribute. In the correlated Gaussian-attributed Erdős-Rényi model, we expect that utilizing node attributes may lead to an efficient algorithm.
\end{itemize}

\section*{Acknowledgment}
This research was supported by the National Research Foundation of Korea under grant 2021R1C1C11008539.

\newpage
\bibliographystyle{IEEEtran}
\bibliography{main}
\newpage
\appendices

\newpage
\appendices
\section{Proof of Lemmas on Erdős-Rényi model}
\begin{IEEEproof}[Proof of Lemma \ref{lem:size of H}]
  The definition of $\mathcal{H}(\pi)$ \eqref{def:H} indicates that $\mathcal{H}_*$ is the set of isolated vertices in the intersection graph $G_1 \wedge_{\pi_*} G_2 $.
  Let us denote $\mathcal{I}$ as the number of isolated vertices in $G\sim \mathcal{G}(n,p)$. Suppose that $np \leq \log n -\omega(1)$.
    Then, we can have
    \begin{equation}
    \begin{aligned}
          \E[\mathcal{I}]&=n(1-p)^{n-1}\\
          &= n\left( 1+ \frac{p}{1-p}\right)^{-n+1}\\
          &\stackrel{(a)}{\geq} n \left( \exp \left( \frac{p}{1-p}\right)\right)^{-n}\\
          &= n \exp\left( -\frac{np}{1-p}\right)\\
          & \geq\frac{1}{2} n^{1- \frac{np}{\log n}} \to \infty.
    \end{aligned}
          \end{equation}
    The inequality $(a)$ holds since $e^x \geq 1+x $, and the last equality holds by assumption $np\leq \log n - \omega(1)$.
Let $\mathcal{I}_i$ be the indicator of the event that $i$ is an isolated vertex. Then, we obtain that
    \begin{equation}
    \begin{aligned}
        &\E[\mathcal{I}^2]=\E[(\Sigma^{n}_{i=1} \mathcal{I}_i)^2]\\
        & = \Sigma^{n}_{i=1}\E[ \mathcal{I}^2_i]+2\Sigma_{i<j}\E[\mathcal{I}_i\mathcal{I}_j]\\
        & = n(1-p)^{n-1}+n(n-1)(1-p)^{2n-3}.
    \end{aligned}       
    \end{equation}
By Chebyshev's inequality, we can obtain that 
    \begin{equation}
    \begin{aligned}
       & \P\left(\mathcal{I}\leq \frac{1}{2}\E[\mathcal{I}] \right) \leq \frac{4\operatorname{Var}[\mathcal{I}]}{\E[\mathcal{I}]^2}\\
        & = \frac{1}{\E[\mathcal{I}]}+\frac{np-1}{n-np}= o(1).
    \end{aligned}
    \end{equation}
    Therefore, $\mathcal{I} \geq \frac{1}{2}\E[\mathcal{I}]\geq \frac{1}{4} n^{1-\frac{np}{\log n}}$ with probability $1-o(1)$. This means that $|\mathcal{H}_*|\geq \frac{1}{4}n^{1-\frac{np_{11}}{\log n}}$ with probability $1-o(1)$ since $G_1 \wedge_{\pi_*} G_2  \sim \mathcal{G}(n,p_{11})$. 
\end{IEEEproof}

\begin{IEEEproof}[Proof of Lemma \ref{lem:mu}]
    By definition of $\mathcal{T}_*$ (Definition \ref{def:T}), we have $\pi(i)=\pi_*(i)$ and $\pi(j)=\pi_*(j)$ for all $ i,j \in [n]\backslash \mathcal{H}_*$. Thus, we have that $A_{i,j}B_{\pi(i),\pi(j)}=A_{i,j}B_{\pi_*(i),\pi_*(j)}$ for all $ i,j \in [n]\backslash \mathcal{H}_*$. On the other hand, if $i\in \mathcal{H}_*$ or $j\in \mathcal{H}_*$, then   we obtain $A_{i,j}B_{\pi(i),\pi(j)} \geq A_{i,j}B_{\pi_*(i),\pi_*(j)}$ since $A_{i,j}B_{\pi_*(i),\pi_*(j)}=0$. Thus, $\mu_{11}(\pi)\geq \mu_{11}(\pi_*)$.
\end{IEEEproof}

\section{Proof of Lemma \ref{lem:k-core}}
\subsection{Notation}
For two random variables $Q$ and $W$, if $W$ is stochastically dominated by $Q$, we write it as $W\preceq Q$.

In this section, we consider $(G_1,G_2)\sim \mathcal{CG}(n,\mathbf{p})$.
For a matching $(M,\mu)$, define 
\begin{equation}
    f(M,\mu)=\Sigma_{i\in M : \mu(i)\neq \pi_*(i)} \deg_{G_1 \wedge_{\mu} G_2}(i).
\end{equation}
The weak $k$-core matching and $\pi_*$-maximal matching are defined in \cite{CKNP20,GRS22,RS23}. For the completeness, we define those again.
\begin{definition}[Weak $k$-core matching]
    We say that a matching $(M,\mu)$ is weak $k$-core matching if 
    \begin{equation}
        f(M,\mu)\geq k |\{i\in M : \mu(i)\neq \pi_*(i)\}|.
    \end{equation}
    
    \end{definition}

\begin{definition}[$\pi_*$-maximal matching]
    We say that a matching $(M,\mu)$ is $\pi_*$-maximal if for every $i \in [n]$, either $i\in M$ or $\pi_*(i) \in \mu(M)$, where $\mu(M)$ is the image of $M$ under $\mu$. Additionally, let us define $\mathcal{M}(d)$ as the set of $\pi_*$-maximal matchings that have $d$ errors. It means that
    \begin{equation}
    \begin{aligned}
         \mathcal{M}(d):=\{(M,\mu) : (M,\mu) \text{ is } \pi_* \text{-maximal and } \\|{i\in M : \mu(i)\neq \pi_*(i)}| =d \}
    \end{aligned}       
    \end{equation}
\end{definition}

Recall that $M_k$ is the $k$-core of the graph $G_1 \wedge_{\pi_*} G_2$. The following lemma was proved in \cite{CKNP20,GRS22}. It states a condition for the $k$-core estimator to have no mismatched node pairs.
\begin{lemma}[Lemma 1 in \cite{CKNP20}]\label{lem:k-core lem1}
    Consider the $(G_1,G_2)\sim \mathcal{CG}(n,\mathbf{p})$. For any positive integer $k$, define the quantity
\begin{equation}
    \xi:=\max _{1 \leq d \leq n} \max _{(M, \mu) \in \mathcal{M}(d)} \mathbb{P}(f(M,\mu) \geq k d)^{1 / d} .
\end{equation}
Then, the $k$-core estimator $(\widehat{M}_k, \widehat{\mu}_k)$ satisfies that
\begin{equation}
    \begin{aligned}
\mathbb{P}\left(\widehat{M}_k=M_k \text { and } \widehat{\mu}_k\{\widehat{M}_k\}\right. & \left.=\pi_*\{\widehat{M}_k\}\right) \geq 2-\exp \left(n^2 \xi\right).
\end{aligned}
\end{equation}
\end{lemma}

The next lemma represents a lower bound on $\xi$. It has been proven by Gaudio et al. \cite{GRS22} in the subsampling model, and we obtained similar results in the correlated ER model.
\begin{lemma}\label{lem:k-core lem2}
 Consider the $(G_1,G_2)\sim \mathcal{CG}(n,\mathbf{p})$.
    For any matching $(M, \mu) \in \mathcal{M}(d)$ and any $\theta>0$, we have that
\begin{equation}
\begin{aligned}
    \mathbb{P}(f(M,\mu) & \geq k d)  \leq 3 \exp \left(-d\left(\theta k-e^{2 \theta} p_{11}-n e^{6 \theta} p_{1*}p_{*1}\right)\right).
\end{aligned}
\end{equation}
\end{lemma}

\begin{IEEEproof}[Proof of Lemma \ref{lem:k-core lem2}]
        Our proof is analogous to the proof of Lemma 23 in \cite{GRS22}. Define the following sets:
    \begin{equation}
  \begin{aligned}
& \mathcal{A}(\mu):=\left\{(i, j) \in[n]^2: i \in M \text { and } \mu(i) \neq \pi_*(i)\right\}  \\
& \mathcal{B}(\mu):=\left\{(i, j) \in[n]^2: \mu(i)=\pi_*(j) \text { and } \mu(j)=\pi_*(i)\right\} \\
& \mathcal{C}(\mu):=\mathcal{A}(\mu) \backslash \mathcal{B}(\mu) .
\end{aligned}
    \end{equation}
Since $(M,\mu) \in \mathcal{M}(d)$, we obtain that $|\mathcal{A}(\mu)|\leq dn$ and $|\mathcal{B}(\mu)|\leq d$. We obtain that 
\begin{equation}
    \begin{aligned}
&f(M,\mu) \\& =\sum_{i \in M: \mu(i) \neq \pi_*(i)} \operatorname{deg}_{G_1 \wedge_\mu G_2}(i)=\sum_{(i, j) \in \mathcal{A}(\mu)} A_{i j} B_{\mu(i) \mu(j)} \\
& =\sum_{(i, j) \in \mathcal{B}(\mu)} A_{i, j} B_{\mu(i), \mu(j)}+\sum_{(i, j) \in \mathcal{C}(\mu)} A_{i, j} B_{\mu(i) ,\mu(j)} \\
& =2 \sum_{(i, j) \in \mathcal{B}(\mu): i<j} A_{i, j} B_{\mu(i) ,\mu(j)}+\sum_{(i, j) \in \mathcal{C}(\mu)} A_{i, j} B_{\mu(i) ,\mu(j)} .
\end{aligned}
\end{equation}
Let $X_{\mathcal{B}} = \sum_{(i, j) \in \mathcal{B}(\mu): i<j} A_{i, j} B_{\mu(i) ,\mu(j)}$ and $X_{\mathcal{C}}=\sum_{(i, j) \in \mathcal{C}(\mu)} A_{i, j} B_{\mu(i) ,\mu(j)}$. Then we obtain that 
\begin{equation}\label{eq:XB}
    X_{\mathcal{B}} \preceq \operatorname{Bin}(d,p_{11})
\end{equation}
since $|\mathcal{B}(\mu)|\leq d$ and $A_{i,j}B_{\mu(i),\mu(j)}=A_{i,j}B_{\pi_*(i),\pi_*(j)}$ for $(i,j)\in \mathcal{B}(\mu)$. Let us construct graph $H$ with vertex set $\{\{i,j\} : (i,j) \in \mathcal{C}(\mu)\}$, where an edge is generated between two vertices $\{i,j\}$ and $\{u,v\}$, if and only if $\{\mu(i), \mu(j)\}=\left\{\pi_*(u), \pi_*(v)\right\}$ or $\{\mu(u), \mu(v)\}=\left\{\pi_*(i), \pi_*(j)\right\}$. Since all vertices in the graph $H$ can have at most two neighbors, the graph $H$ is 3-colorable. Let us denote the partitions formed by each color as $\mathcal{C}_1$, $\mathcal{C}_2$, and $\mathcal{C}_3$. For any $\{i,j\}, \{u,v\} \in \mathcal{C}_1$, there is no edge between $\{i,j\}$ and $ \{u,v\}$, indicating that $\{\mu(i), \mu(j)\}\neq\left\{\pi_*(u), \pi_*(v)\right\}$ and $\{\mu(u), \mu(v)\}\neq \left\{\pi_*(i), \pi_*(j)\right\}$. Therefore, $A_{i,j}B_{\mu(i),\mu(j)}$ and $A_{u,v}B_{\mu(u),\mu(v)}$ are independent. Hence, we obtain that
\begin{equation}\label{eq:XC1}
\begin{aligned}
X_{\mathcal{C}_1}:=\sum_{\{i, j\} \in \mathcal{C}_1} A_{i, j} B_{\mu(i), \mu(j)} \preceq \operatorname{Bin}\left(dn,p_{1*}p_{*1}\right)
\end{aligned}
\end{equation}
since $|\mathcal{C}_1| \leq |\mathcal{C}|\leq dn$ and $A_{i, j} B_{\mu(i), \mu(j)}\sim \operatorname{Bern}(p_{1*}p_{*1})$ for $(i,j)\in \mathcal{C}$.
Similarly, we can obtain that
\begin{equation}\label{eq:XC23}
    \begin{aligned}
X_{\mathcal{C}_2},X_{\mathcal{C}_3} \preceq \operatorname{Bin}\left(dn,p_{1*}p_{*1}\right).
\end{aligned}
\end{equation}
Finally, we can have that
\begin{equation}
    \begin{aligned}
& \mathbb{P}(f(M,\mu) \geq k d) \leq \mathbb{P}\left(2 X_{\mathcal{B}}+X_{\mathcal{C}} \geq k d\right) \\
& \leq \mathbb{P}\left(2\left(X_{\mathcal{B}}+X_{\mathcal{C}_1}+X_{\mathcal{C}_2}+X_{\mathcal{C}_3}\right) \geq k d\right) \\
& \leq \sum_{i=1}^3 \mathbb{P}\left(2 X_{\mathcal{B}}+6 X_{\mathcal{C}_i} \geq k d\right) \\
&\stackrel{(a)}{\leq} 3 e^{-\theta k d} \mathbb{E}\left[e^{2 \theta X_{\mathcal{B}}}\right] \mathbb{E}\left[e^{6 \theta X_{\mathcal{C}_i}}\right] \\
& \stackrel{(b)}{\leq} 3 e^{-\theta k d}\left(1+p_{11}\left(e^{2 \theta}-1\right)\right)^d\left(1+p_{1*}p_{*1}\left(e^{6 \theta}-1\right)\right)^{d n} \\
& \leq \exp \left\{-d\left(\theta k-e^{2 \theta} p_{11 }-n e^{6 \theta} p_{1*}p_{*1}\right)\right\} .
\end{aligned}
\end{equation}
The inequality $(a)$ holds from the Chernoff bound. The inequality $(b)$ holds by combining \eqref{eq:XC1}, \eqref{eq:XC23}, and the moment generating function of the binomial random variable. The last inequality holds by $e^x \geq 1+x$. Thus, the proof is complete.    
\end{IEEEproof}

\begin{IEEEproof}[Proof of Lemma \ref{lem:k-core}]
    By Lemma \ref{lem:k-core lem1}, if we show that $\xi=o(n^{-2})$, then we can complete the proof.
    By Lemma \ref{lem:k-core lem2}, for any $\theta>0$, we obtain that
    \begin{equation}
        \begin{aligned}
            \xi &\leq 3^{1/d} \exp \left(-\theta k+e^{2 \theta} p_{11}+n e^{6 \theta} p_{01}p_{10}\right)\\
            & \leq \exp \left(-\theta k+e^{2 \theta} p_{11}+n e^{6 \theta}  p_{1*}p_{*1}\right).
        \end{aligned}
    \end{equation}
    Therefore, it is sufficient to show the existence of some $\theta$ such that $\theta k-e^{2 \theta} p_{11}-n e^{6 \theta}  p_{1*}p_{*1} \geq 2\log n +\omega(1)$. Let $\theta=(\log \log n)^{2.5}$. By assumption \eqref{eq:k-core k regmie}, we obtain that 
    \begin{equation}\label{eq:1}
        \theta k =\omega(\log n).
    \end{equation}
    We can have that
    \begin{equation}\label{eq:2}
        e^{2\theta}p_{11}=o(1)
    \end{equation}
    since the assumption \eqref{eq:k-core p regmie} implies that $p_{11}\leq O\left(e^{-(\log \log n)^3}\right)$. We also obtain that 
    \begin{equation}\label{eq:3}
        \begin{aligned}
            \theta k &\stackrel{(a)}{\geq} \theta \frac{np_{11}}{(\log np_{11})^2} \stackrel{(b)}{\geq} \frac{np_{11}}{(\log n)^2} \stackrel{(c)}{\geq} 2ne^{6\theta}p_{1*}p_{*1}.
        \end{aligned}
    \end{equation}
    The inequality $(a)$ holds by assumption \eqref{eq:k-core k regmie}, the inequality $(b)$ holds since $\theta=(\log \log n)^{2.5}$ and $p_{11}\leq 1$, and the inequality $(c)$ holds since assumption \eqref{eq:k-core p regmie}.
    Combining \eqref{eq:1}, \eqref{eq:2} and \eqref{eq:3} yields that  $\theta k-e^{2 \theta} p_{11}-n e^{6 \theta}  p_{1*}p_{*1} \geq 2\log n +\omega(1)$. Hence, the proof is complete.
\end{IEEEproof}

\section{Proof of Lemmas regarding set size}
Recall that  $  L_k := \{i\in [n] : \deg_G(i) \leq k\}$. 
\begin{IEEEproof}[Proof of Lemma \ref{lem:size degree k}]
    We obtain that
    \begin{equation}
        \begin{aligned}
              \P(\deg_G(i) \leq k)&=\P(\operatorname{Bin}(n,p)\leq k)\\
            & \stackrel{(a)}{\leq} \inf_{t>0}(1-p+pe^{-t})^n e^{kt}\\
            &\stackrel{(b)}{\leq} \inf_{t>0} \exp\left(-np(1-e^{-t})+kt \right)\\
            &\leq \exp\left(-np + k \log np+1\right).
        \end{aligned}
    \end{equation}
    The inequality $(a)$ holds by Chernoff bound. The inequality $(b)$ holds since $p(1-e^{-t})>0$ for $t>0$ and the last inequality holds by choosing $t= \log np$.
\end{IEEEproof}

Let $F_k$ be the set of vertices outside the $k$-core of the graph $G$.
To prove Lemma \ref{lem:3L}, we will use lemma proven in \cite{GRS22}.
\begin{lemma}[Lemma IV.6 in \cite{GRS22}]\label{lem:IV}
    Let $G\sim \mathcal{G}(n,p)$. Suppose that $p\leq \gamma/n$ with $\gamma = o(\sqrt{n})$. If $|L_{k+1}|\leq \frac{n}{4\gamma^2}$ then we obtain that $|F_k|\leq 3|L_{k+1}|$.
\end{lemma}
\begin{IEEEproof}[Proof of Lemma \ref{lem:3L}]
    In the proof of Theorem \ref{thm:k-core matching},  we showed that
\begin{equation}
    |\hat{L}_{k+1}| \leq n^{1-\frac{np_{11}-(k+1) \log np_{11}}{\log n} +o(1)}
\end{equation}
with probability $1-o(1).$ By the assumption $p_{11}=\Theta\left(\frac{\log n}{n}\right)$, we can obtain $ |\hat{L}_{k+1}| \leq \frac{1}{4np^2_{11}}$. Thus, we can conclude the proof by applying Lemma \ref{lem:IV}.
\end{IEEEproof}

\end{document}